\documentclass[prl,aps,twocolumn,floatfix,showpacs]{revtex4-1}
\usepackage{graphicx,amsmath,amssymb,times}

\begin{document}
\title{Route to supersolidity for the extended Bose-Hubbard model}
\author{M. Iskin}
\affiliation{Department of Physics, Ko\c c University, Rumelifeneri Yolu, 34450 Sariyer, Istanbul, Turkey}
\date{\today}

\begin{abstract}
We use the Gutzwiller ansatz and analyze the phase diagram 
of the extended Bose-Hubbard Hamiltonian with on-site ($U$) and 
nearest-neighbor ($V$) repulsions. For $d$-dimensional hypercubic 
lattices, when $2dV < U$, it is well-known that the ground state 
alternates between the charge-density-wave (CDW) 
and Mott insulators, and the supersolid (SS) phase occupies small 
regions around the CDW insulators. However, when $2dV > U$, in this paper,
we show that the ground state has only CDW insulators, 
and more importantly, the SS phase occupies a much larger region 
in the phase diagram, existing up to very large hopping values 
which could be orders of magnitude higher than that of the well-known
case. In particular, the SS-superfluid phase boundary increases 
linearly as a function of hopping when $2dV \gtrsim 1.5U$, for which 
the prospects of observing the SS phase with dipolar Bose gases 
loaded into optical lattices is much higher.
\end{abstract}

\maketitle

Can a solid be superfluid? The so-called \textit{supersolid} 
phase is characterized by the simultaneous existence of diagonal
(crystalline) and off-diagonal (superfluid) long-range orders. 
Although this intriguing possibility was suggested a long time ago in 
the context of solid $^4$He~\cite{supersolid}, and in spite 
of numerous attempts over the past decades, a convincing 
experimental evidence for its existence is yet to be found~\cite{leggett}. 
On one hand, there is still some controversy in 
the condensed matter literature around the recent reports that the 
theoretically predicted nonclassical rotational inertia was found 
in solid $^4$He with the torsional oscillator experiments~\cite{He4}. 
On the other hand, there is strong theoretical evidence that 
the situation in lattice models is promising~\cite{goral,sengupta}, 
which could be advanced with ultracold quantum gases loaded 
into optical lattices~\cite{bloch}.

Possibly the simplest models that show SS behavior are the extended-type 
Bose-Hubbard ones with on-site and nearest-neighbor (NN) repulsions.
Since these models can be naturally realized with dipolar bosons~\cite{dipolar}, 
i.e. bosonic atoms or molecules with permanent or induced magnetic or 
electric dipole moments, the ground-state phase diagrams of various 
extended models have already been studied.
For instance, the existence and stability of SS phases have been 
demonstrated via the Gutzwiller ansatz~\cite{goral, kovrizhin, menotti} and 
decoupling mean-field~\cite{mf} approaches, and numerically exact 
quantum Monte Carlo~\cite{sengupta, qmc} techniques.
The SS is known to be a very fragile phase, and one of the major obstacles 
in creating and observing it is its very small existence region and 
low critical temperatures. In a very recent proposal~\cite{buhler}, 
it has been suggested that one way of overcoming these difficulties is
to load high occupancies of bosons into optical lattices.
Such systems are well-described by the quantum rotor model, for which
the mean-field calculation gives linear dependence between the critical
hopping ($t_c$) and occupancy.

In this paper, we use the Gutzwiller ansatz and mean-field 
decoupling to analyze the phase diagram of the extended 
Bose-Hubbard Hamiltonian with on-site ($U$) and isotropic NN 
($V$) repulsions. For $d$-dimensional hypercubic lattices, 
beyond the critical threshold $2dV > U$, we show that the SS phase 
occupies a much larger region in the phase diagram, existing up to 
very large hopping values of the order $2dt_c \gtrsim U$. 
In particular, the SS-superfluid phase boundary increases 
linearly as a function of hopping when $2dV \gtrsim 1.5U$.
This must me contrasted with the below threshold ($2dV < U$) case, 
for which it is well-known that the SS phase occupies small regions 
around the CDW insulators, existing only up to 
$2dt_c \lesssim 0.4U$. Therefore, we show that the prospects of 
observing the SS phase is much higher when $2dV > U$. 
We also argue that our results for two dimensions is directly 
applicable to the quasi-two-dimensional dipolar Bose gases, 
for which the condition $4V > U$ could be easily achieved by tuning 
the $s$-wave scattering length using the Feshbach resonances.

\textit{Hamiltonian}:
To obtain these results, we use the extended Bose-Hubbard Hamiltonian
with an isotropic NN repulsion
\begin{align}
\label{eqn:ham}
H = &- t\sum_{\langle i,j \rangle} (b_i^\dagger b_j + b_j^\dagger b_i) 
+ \frac{U}{2} \sum_i \widehat{n}_i (\widehat{n}_i-1) \nonumber \\
&+ V \sum_{\langle i,j \rangle} \widehat{n}_i \widehat{n}_j -\mu \sum_i \widehat{n}_i,
\end{align}
where $t$ is the tunneling (or hopping) amplitude between NN sites $i$ 
and $j$, $b_i^\dagger$ ($b_i$) is the boson creation (annihilation) 
operator at site $i$, $\widehat{n}_i = b_i^\dagger b_i$ is the boson 
number operator, and $\mu$ is the chemical potential. 
As it turns out, the phase diagram of this Hamiltonian depends strongly on 
the relative strength $U-zV$, where $z = 2d$ is the lattice coordination 
number, leading to two important cases.

\textit{(I) Weak-NN-coupling ($zV < U$)}:
When $V \ne 0$, the ground state has two types of insulating 
phases~\cite{kovrizhin, iskin2}. 
The first one is the Mott insulator where, similar to the usual 
Bose-Hubbard model, the average boson occupancy is the same 
for every lattice site, i.e. $\langle \widehat{n}_i \rangle = n_0$. Here, 
$\langle ... \rangle$ is the thermal average, and $n_0$ is chosen to 
minimize the ground-state energy for a given $\mu$.
The second one is the CDW insulator which has crystalline order in 
the form of staggered average occupancies. To describe the CDW insulator, 
it is convenient to split the entire lattice into two sublattices 
(e.g. $A$ and $B$) such that the NN sites belong to a different 
sublattice, i.e. $\langle \widehat{n}_i \rangle = n_A$ and 
$\langle \widehat{n}_j \rangle = n_B$ for $\langle i, j \rangle$. 
We assume the occupancies are such that $n_A \ge n_B$, and the 
case with $n_A = n_B$ corresponds to the Mott insulator.

\begin{figure} [htb]
\centerline{\scalebox{0.35}{
\includegraphics{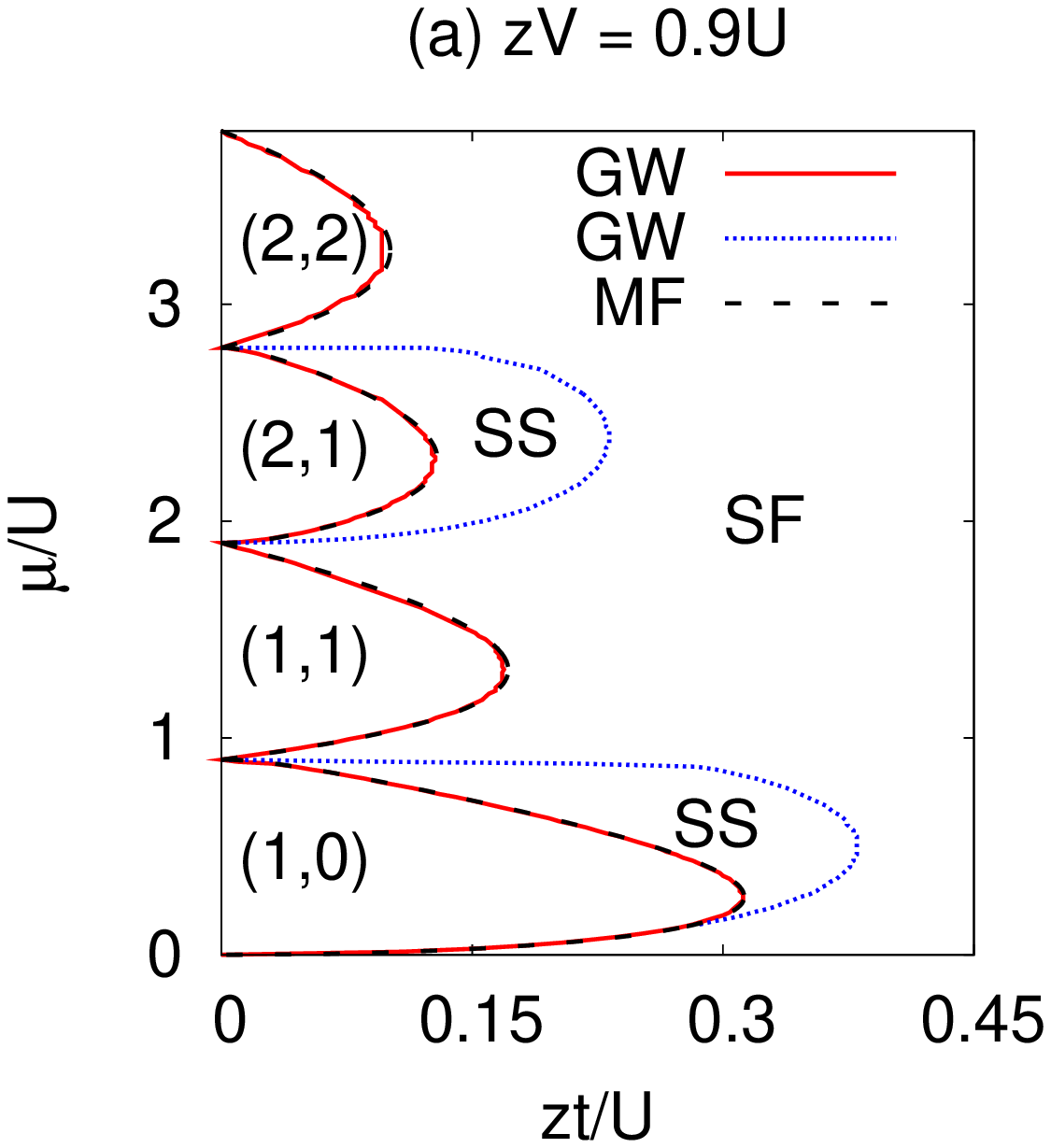} 
\includegraphics{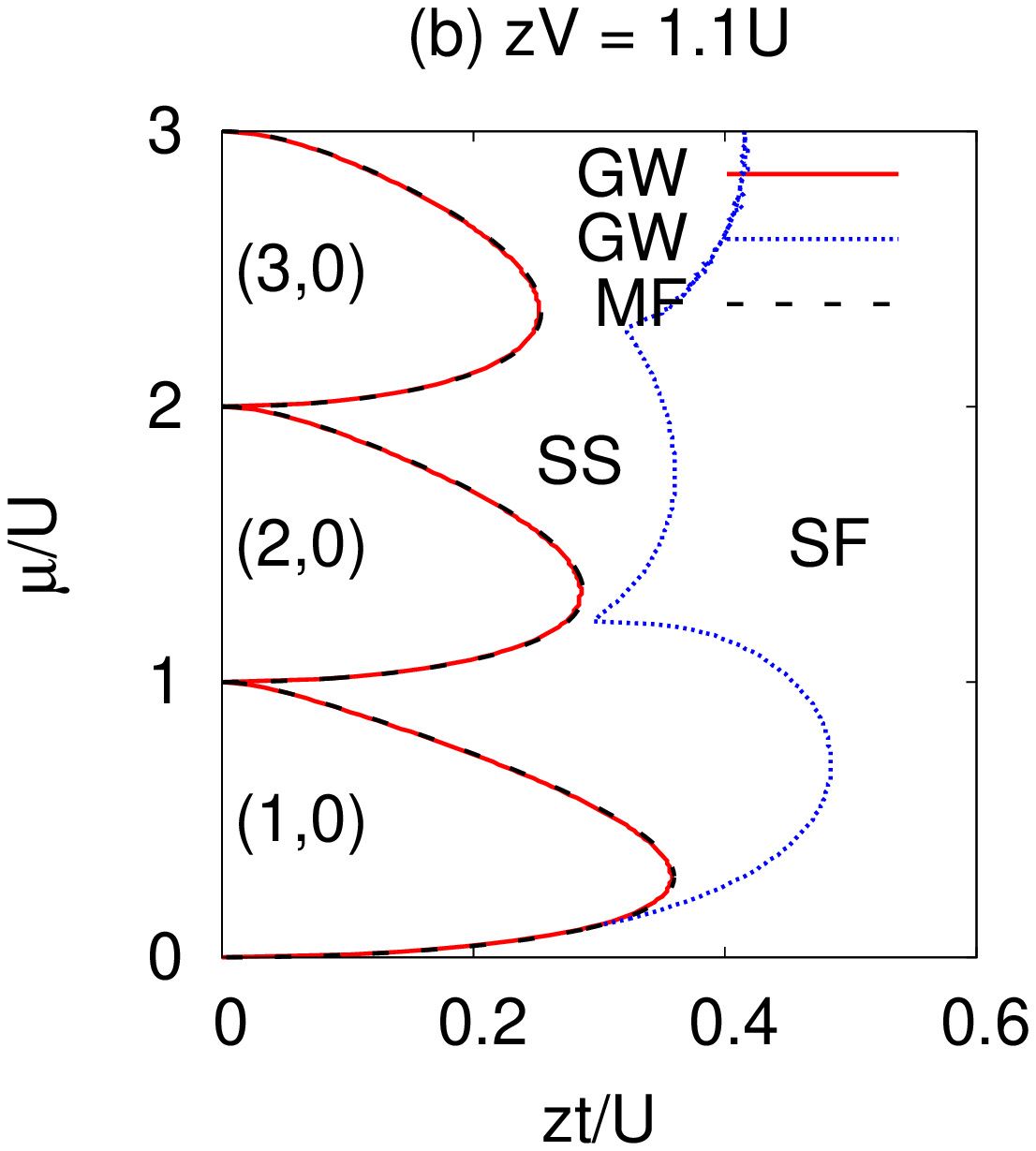}}}
\centerline{\scalebox{0.35}{
\includegraphics{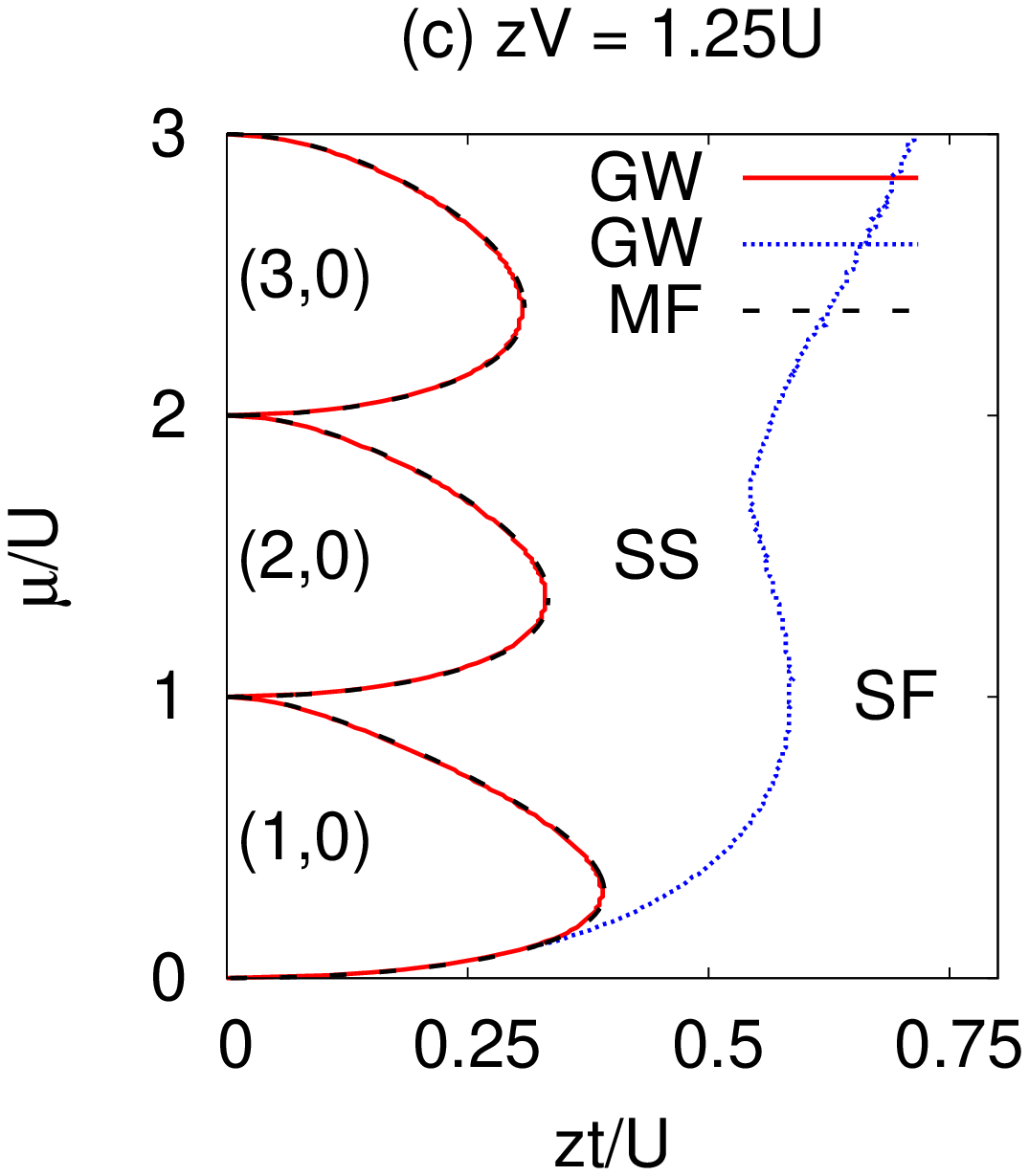} 
\includegraphics{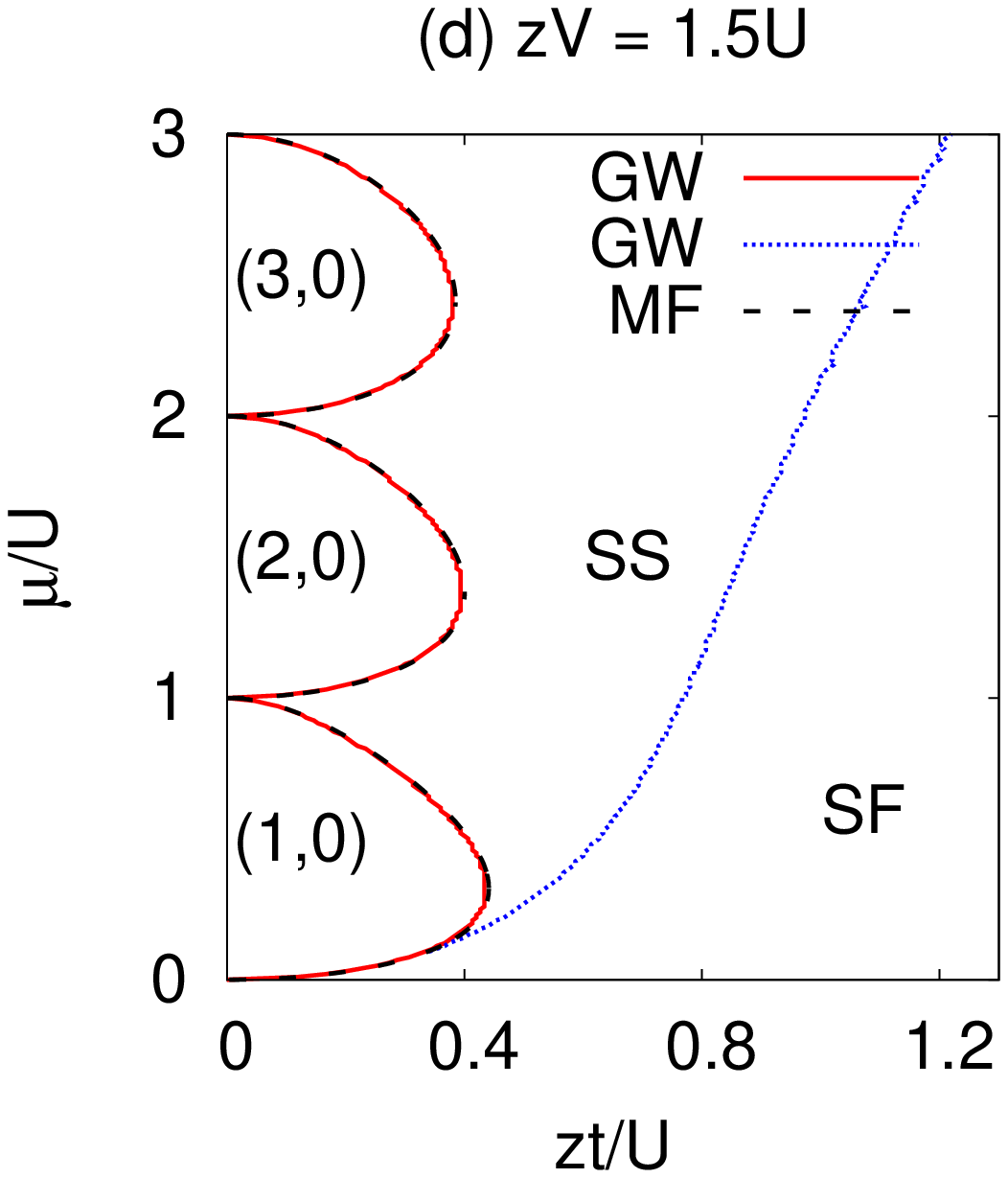}}}
\centerline{\scalebox{0.35}{
\includegraphics{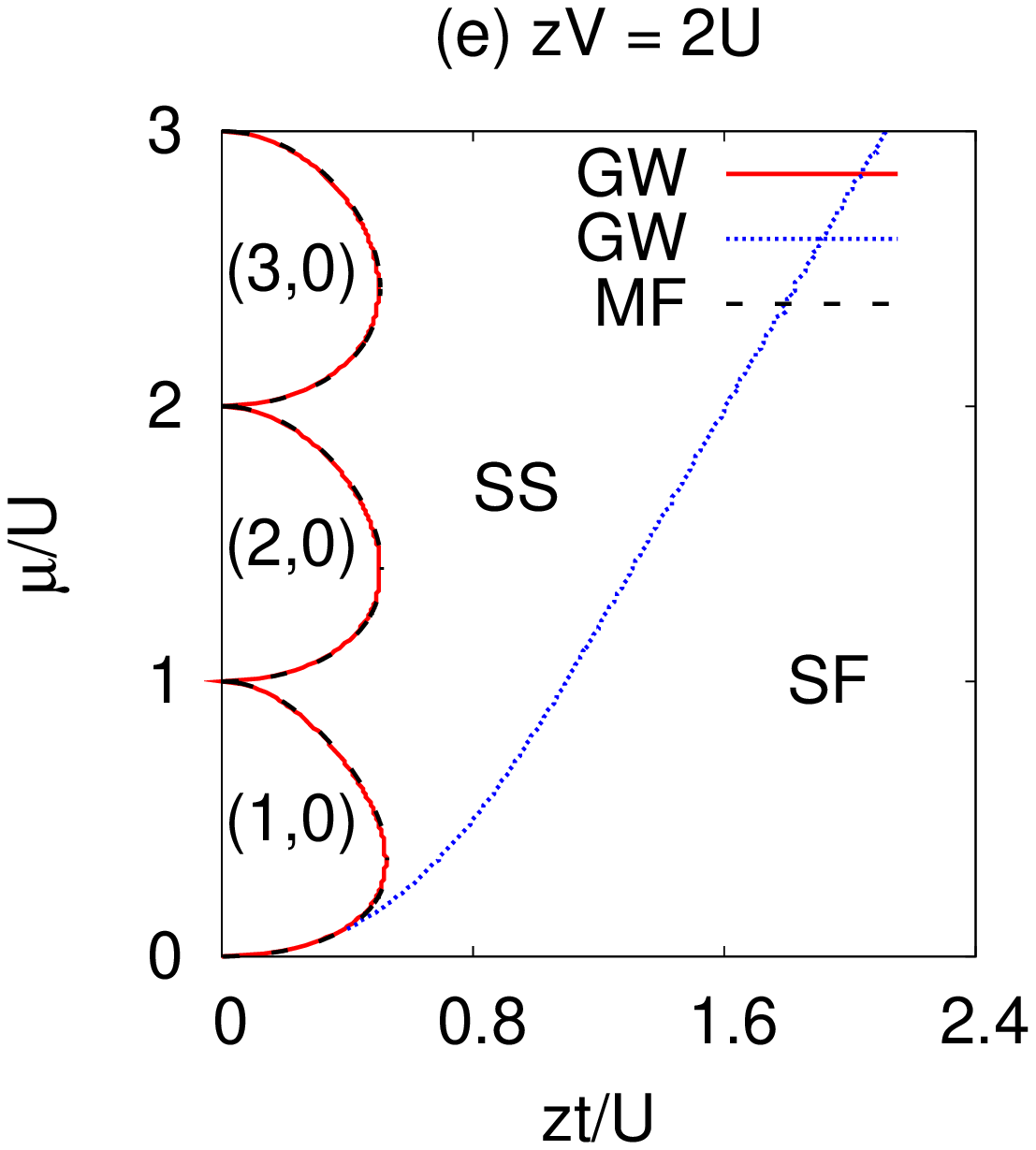} 
\includegraphics{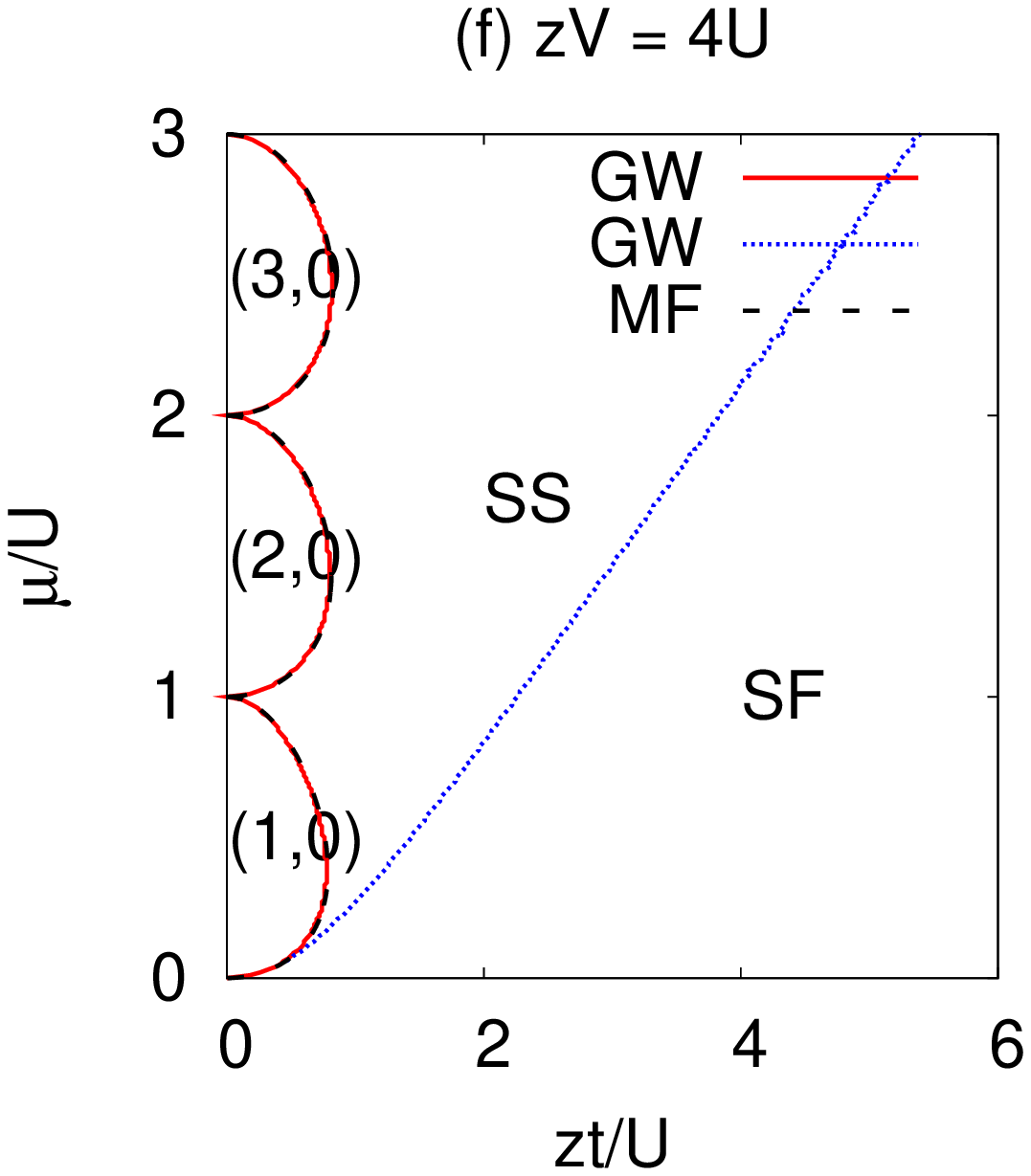}}}
\caption{\label{fig:pd} (Color online)
The ground-state phase diagrams are shown as a function of the chemical 
potential $\mu$ and hopping $t$ for
(a) $zV = 0.9U$, (b) $zV = 1.1U$, (c) $zV = 1.25U$,
(d) $zV = 1.5U$, (e) $zV = 2U$, and (f) $zV = 4U$.
The CDW and Mott insulators are indicated with their sublattice 
occupancies $(n_A,n_B)$. Here, the red continuous and blue dotted 
lines are obtained from the Gutzwiller ansatz calculations, and 
the black dashed lines are obtained from Eq.~(\ref{eqn:mf-cdw}). 
}
\end{figure}

In the atomic ($t = 0$) limit, it turns out that the chemical 
potential width of all CDW and Mott insulators are $zV$ and $U$, 
respectively, and that the ground state alternates between the 
CDW and Mott phases as a function of increasing $\mu$. 
For instance, the ground state 
   is a $(n_A = 0, n_B = 0)$ vacuum for $\mu \le 0$; 
a $(1, 0)$ CDW insulator for $0 < \mu < zV$; 
a $(1, 1)$ Mott insulator for $zV < \mu < U+zV$; 
a $(2, 1)$ CDW insulator for $U+zV < \mu < U+2zV$; 
a $(2, 2)$ Mott insulator for $U+2zV < \mu < 2U+2zV$, and so on.
As $t$ increases, the range of $\mu$ about which the ground state 
is insulating decreases, and the CDW and Mott insulators disappear at 
a critical value of $t$, beyond which the system becomes compressible 
(SF or SS) as shown in Fig.~\ref{fig:pd}(a).

\textit{(II) Strong-NN-coupling ($zV > U$)}:
In contrast to the well-known weak-NN-coupling, the strong-NN-coupling  
of this model has not been studied much in the literature, 
which is the main topic of this paper. 
When $zV$ equals exactly to $U$, it is easy to check at least in the 
atomic limit that the $(n_0+1, n_0)$ CDW insulator becomes 
degenerate in energy with the $(2n_0+1, 0)$ CDW insulator, 
and the $(n_0, n_0)$ Mott insulator becomes
degenerate with the $(2n_0, 0)$ CDW insulator.
This indicates that, beyond the critical $zV = U$ threshold, both the 
CDW and Mott insulators that are found in the weak-NN-coupling are 
unstable against formation of new CDW insulators, leading to a 
phase diagram which has a very different qualitative structure. 

In this paper, we show that the ground state has only CDW-type
insulating phases in the strong-NN-coupling, the chemical potential 
width of all are $U$ in the atomic limit. For instance, the ground state 
   is a $(0, 0)$ vacuum for $\mu \le 0$,
a $(1, 0)$ CDW insulator for $0 < \mu < U$,
a $(2, 0)$ CDW insulator for $U < \mu < 2U$,
a $(3, 0)$ CDW insulator for $2U < \mu < 3U$, and so on.
As $t$ increases, the range of $\mu$ about which the ground state 
is insulating decreases, and the CDW insulators disappear at a critical 
hopping, beyond which the system becomes a SS, as shown in Figs.~\ref{fig:pd}(b-f).
Most importantly, unlike the weak-NN-coupling where the SS phase 
occupies small regions in the phase diagram around the CDW insulators, 
we show that the SS phase occupies a much larger region in this case, 
existing up to very large hopping values which could be orders of 
magnitude higher than that of the weak-NN-coupling. To obtain these 
results, we solve the Schr\"odinger equation for the Gutzwiller ansatz, 
as discussed next.

\textit{Gutzwiller ansatz}:
This ansatz has been frequently used to approximate the many-body wave 
functions of Bose-Hubbard Hamiltonians~\cite{goral, kovrizhin, menotti}.
It can be written as
\begin{equation}
\label{eqn:GW}
| \psi \rangle = \prod_i \left( \sum_m f_{i,m} | i,m \rangle \right),
\end{equation}
where $| i,m \rangle$ represents the Fock state of $m$ bosons 
occupying the site $i$, and $f_{i,m}$ is the probability amplitude
of its occupation. Here, $m = 0, 1, \dots, m_\textrm{max}$, where 
$m_\textrm{max}$ is the maximum number of bosons allowed in the
numerics which we typically choose $m_\textrm{max}=50$.
The normalization of the wave function $\langle \psi | \psi \rangle$ 
requires $\sum_m |f_{i,m}|^2 = 1$ for each $i$. 

Within this ansatz, the superfluid order parameter 
$
\phi_i = \langle \psi | b_i | \psi \rangle 
$ 
is determined by
\begin{equation}
\label{eqn:op}
\phi_i = \sum_m \sqrt{m+1} f_{i,m}^* f_{i,m+1}.
\end{equation}
This complex parameter describes the state of the system at site $i$:
while it vanishes for the CDW and Mott insulators, it is finite for 
the SF and SS ground states. Therefore, $\phi_i \to 0^+$ signals the 
phase boundary between an insulating and a compressible phase. 
Similarly, the average occupancy 
$
n_i = \langle \psi | b_i^\dagger b_i | \psi \rangle
$ 
is determined by
\begin{align}
\label{eqn:num}
n_i = \sum_m m |f_{i,m}|^2.
\end{align}
In Eqs.~(\ref{eqn:op}) and~(\ref{eqn:num}), the probability amplitudes
are obtained by solving the Schr\"odinger equation,
$
\langle \psi | H | \psi \rangle =  i\hbar \langle \psi | \partial | \psi \rangle/\partial \tau,
$
with
$
f_{i,m} = f_{0,i,m} e^{-i \epsilon_{0,i} \tau/\hbar}.
$
This leads to
\begin{align}
\label{eqn:SE}
\epsilon_{0,i} f_{0,i,m} &= -t\left( \bar{\phi}_i \sqrt{m} f_{0,i,m-1} 
+ \bar{\phi}_i^* \sqrt{m+1} f_{0,i,m+1} \right) \nonumber \\
&+ \left[ \frac{U}{2} m (m-1) + V m \bar{n}_i - \mu m \right] f_{0,i,m},
\end{align}
where $\bar{\phi}_i = \sum_{{\langle j \rangle}_i} \phi_j$ and 
$\bar{n}_i = \sum_{{\langle j \rangle}_i} n_j$ sum over sites 
$j$ neighboring to site $i$. 

For the ground state, first we need the minimal eigenvalue $\epsilon_{0,i}$ 
and the elements $f_{0,i,m}$ of the corresponding eigenvector,
and then use them in Eqs.~(\ref{eqn:op}) and~(\ref{eqn:num}) to solve 
for $\phi_i$ and $n_i$ self-consistently.
Note that, in uniform systems with two sublattices discussed in this paper, 
the state of the whole system is sufficiently described by two order parameters: 
$\phi_A$ and $\phi_B$ for the sublattices A and B, respectively.
When this is the case, note also that $\epsilon_{0,i} = \epsilon_{0,A}$, 
$f_{0,i,m} = f_{0,A,m}$, $\bar{\phi}_i = z\phi_B$, $n_i = n_A$ and 
$\bar{n}_i = z n_B$ for $i \in$ A sublattice, and 
$\epsilon_{0,i} = \epsilon_{0,B}$, $f_{0,i,m} = f_{0,B,m}$, 
$\bar{\phi}_i = z\phi_A$, $n_i = n_B$ and $\bar{n}_i = z n_A$ for $i \in$ B sublattice.
To support our Gutzwiller ansatz calculations, next we examine
the mean-field theory, which provides an analytical expression for the
phase boundary between the insulating and compressible phases.

\textit{Mean-field decoupling approximation}:
In constructing the mean-field theory, one first defines the SF order 
parameter $\phi_i = \langle b_i \rangle$, and then replaces the 
operator $b_i$ with $\phi_i + \delta b_i$ in the hopping terms of 
Eq.~(\ref{eqn:ham}). This approximation decouples the two-particle 
hopping terms into single-particle ones, and the resultant 
mean-field Hamiltonian can be solved via exact diagonalization 
in a power series of $\phi_i$.
Performing a second-order perturbation theory in $\phi_i$ around 
the insulators, and following the usual Landau procedure for 
second-order phase transitions, i.e. minimizing the energy as a 
function of $\phi_i$, we eventually arrive at the condition
\begin{equation}
\label{eqn:mf-op}
\phi_i = \bar{\phi}_i t \left[\frac{n_i+1}{U n_i + V \bar{n}_i -\mu} 
- \frac{n_i}{U (n_i-1) + V\bar{n}_i - \mu} \right],
\end{equation}
where the definitions of $\bar{\phi}_i$ and $\bar{n}_i$ are the same
as in Eq.~(\ref{eqn:SE}). It is known that the results of the 
mean-field theory coincide with those of the Gutzwiller ansatz,
and that they both become exact when $d \gg 1$~\cite{rokhsar}. 
We emphasize that the mean-field calculations gives a good 
qualitative description of the system, and it becomes progressively 
accurate as the dimensionality and/or the occupancy increases.

For uniform systems with two sublattices, Eq.~(\ref{eqn:mf-op}) gives 
coupled equations for $\phi_A$ and $\phi_B$, which can be solved
to obtain the phase boundary between the insulating (CDW or Mott) 
and compressible (SF or SS) phases. Since $\phi_A, \phi_B \to 0^+$ 
near these boundaries, Eq.~(\ref{eqn:mf-op}) can be satisfied only if~\cite{iskin1}
\begin{align}
\label{eqn:mf-cdw}
\frac{1}{z^2 t^2} &= \left[ \frac{n_A+1}{U n_A + z V n_B - \mu} 
- \frac{n_A}{U (n_A-1) + z V n_B - \mu} \right] \nonumber \\
&\times \left[ \frac{n_B+1}{U n_B + z V n_A -\mu} -\frac{n_B}{U (n_B-1) + z V n_A -\mu} \right],
\end{align}
which gives a quartic equation for $\mu$. An alternative way of deriving 
this equation is the random-phase approximation~\cite{iskin2}. 
Since a simple closed form analytic solution for $\mu$ is not possible, 
we solve Eq.~(\ref{eqn:mf-cdw}) for each of the insulating lobes separately. 
Having discussed the details of the Gutzwiller ansatz and mean-field 
approximation, we are ready to discuss the phase diagrams.

\textit{Phase diagrams}:
We solve Eqs.~(\ref{eqn:op}-\ref{eqn:SE}) self-consistently for the 
order parameters (i.e. $\phi_A$ and $\phi_B$) and average
occupancies (i.e. $n_A$ and $n_B$), and use them to construct the 
phase diagram of the system. 
The CDW and Mott insulators are characterized by $\phi_A = \phi_B = 0$, 
and $n_A \ne n_B$ and $n_A = n_B$, respectively. However, 
the SF and SS phases are characterized by $\phi_A = \phi_B \ne 0$
and $\phi_A \ne \phi_B$, respectively, which naturally leads to
$n_A = n_B$ in the SF and $n_A \ne n_B$ in the SS phase. 
In this paper, we choose $\phi_A$ and $\phi_B$ to be real, since 
we are only interested in the CDW-SS and MI-SF phase boundaries 
which are determined by $\{\phi_A, \phi_B\} \ne 0$,
and the SS-SF phase boundaries which are determined by 
$\phi_A = \phi_B \ne 0$.

\begin{figure} [htb]
\centerline{\scalebox{0.35}{
\includegraphics{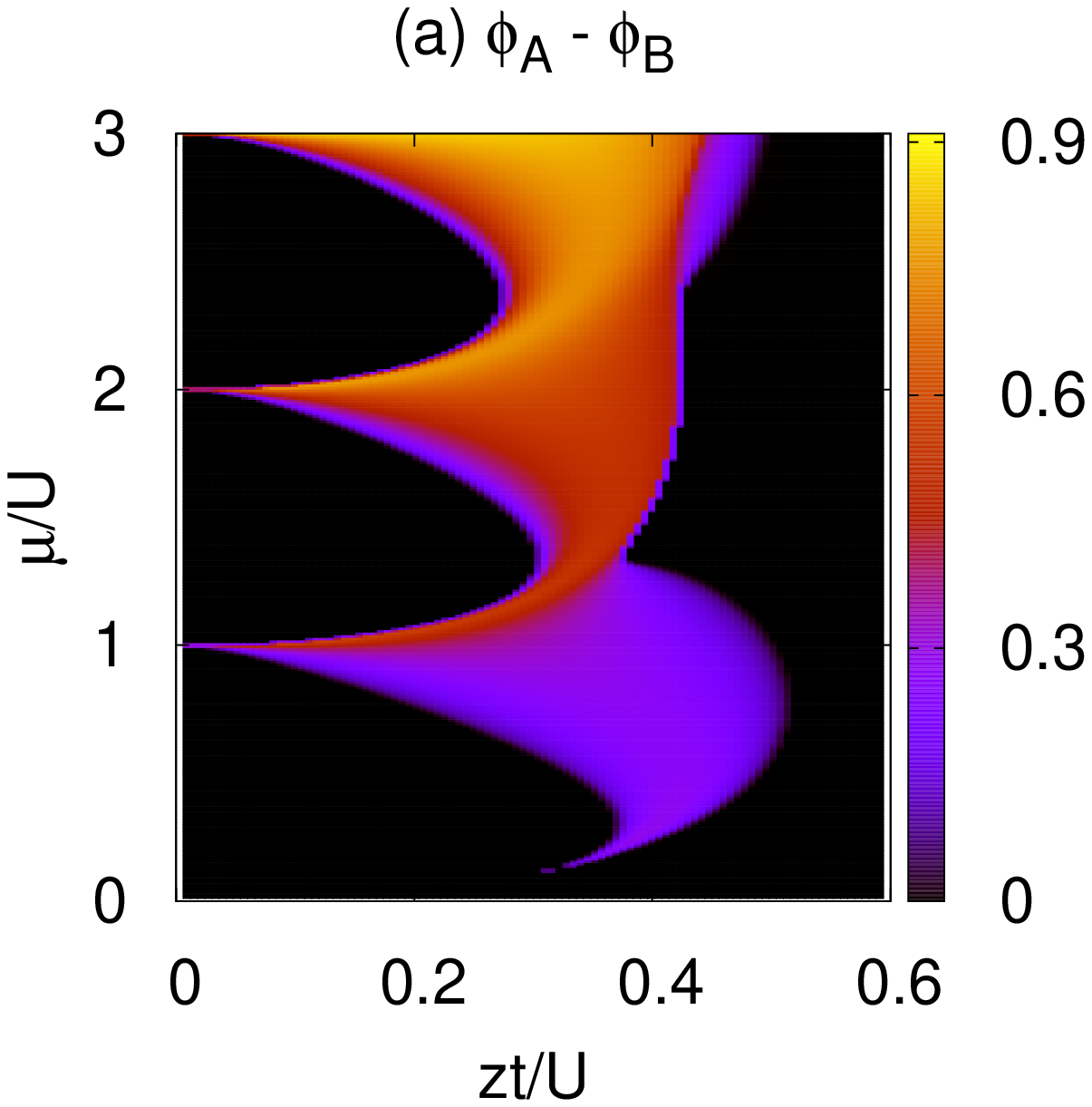} 
\includegraphics{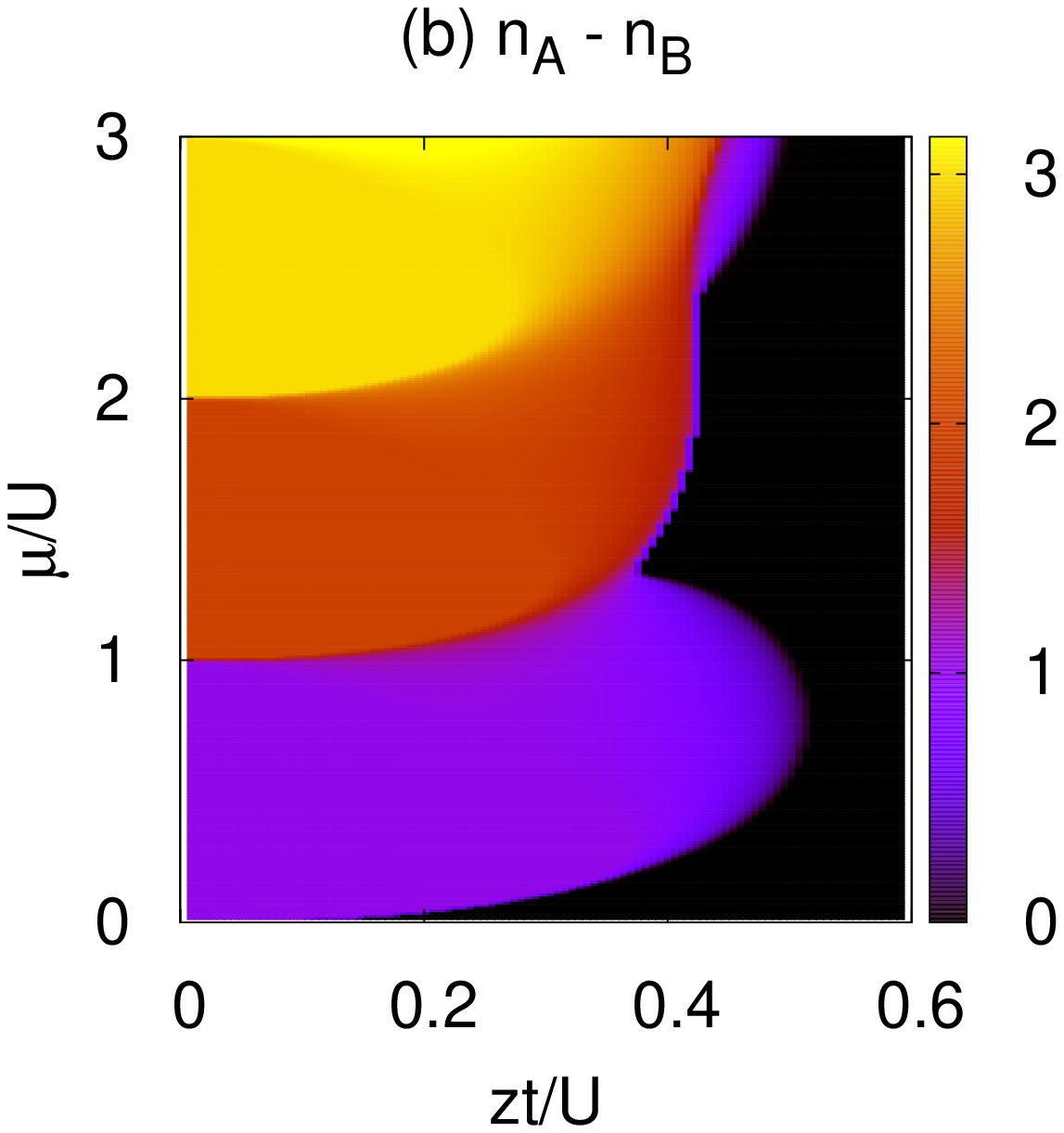}}}
\caption{\label{fig:map} (Color online)
The colored maps of the (a) relative order parameter $\phi_A - \phi_B$ 
and (b) relative average occupancy $n_A - n_B$ are shown as 
a function of the chemical potential $\mu$ and hopping $t$ when $zV = 1.15U$. 
}
\end{figure}

Therefore, it is sufficient to look at the relative order parameter 
$\phi_A - \phi_B$ and relative average occupancy $n_A - n_B$ to 
distinguish between various phases. For instance,
in Fig.~\ref{fig:map}, we show the colored maps of $\phi_A - \phi_B$ 
and $n_A - n_B$ as a function of the chemical potential $\mu$ 
and hopping $t$ for $zV = 1.15U$. The three dark lobes 
shown in Fig.~\ref{fig:map}(a), where $\phi_A = \phi_B = 0$, 
correspond to $(1, 0)$, $(2, 0)$ and $(3, 0)$ CDW insulators 
(from bottom to top), the occupancies of which are clearly seen 
in Fig.~\ref{fig:map}(b). Here, the SF phase occupies the dark region 
that is common in both Figs.~\ref{fig:map}(a) and~\ref{fig:map}(b). 
It is also clear that the SS phase, where $\phi_A - \phi_B \ne 0$ 
and $n_A - n_B \ne 0$ shown with bright colors in Figs.~\ref{fig:map}(a) 
and~\ref{fig:map}(b), respectively, is sandwiched between the 
CDW insulators from the left and SF phase from the right.
Note that, in the SS phase, the superfluid order parameter is larger
on the sublattice with higher occupancy, and both the crystalline
and superfluid orders are primarily on the same sublattice~\cite{sublattice}.
This is because the particle and hole excitation energies are higher on
the sublattice with lower occupancy when $zV > U$.

In Fig.~\ref{fig:pd}, we repeat this analysis for a number of NN 
repulsions, and plot the phase diagrams as a function of $\mu$ and $t$. 
In these figures, the red continuous and blue dotted lines are 
obtained from the Gutzwiller ansatz calculations, and 
the black dashed lines are obtained from Eq.~(\ref{eqn:mf-cdw}). 
Note that both methods are in complete agreement, i.e. on top of 
each other, for the phase boundary between the insulating 
(CDW or Mott) and compressible (SF or SS) phases, which supports our 
Gutzwiller ansatz calculations.

A typical weak-NN-coupling phase diagram is shown in 
Fig.~\ref{fig:pd}(a) for $zV = 0.9U$. As discussed above, 
the ground state alternates between the CDW 
and Mott insulators as a function of increasing $\mu$, and the SS 
phase occupies only small regions around the CDW insulators. 
Unlike the weak-NN-coupling phase diagrams, Figs.~\ref{fig:pd}(b-f) 
show that the ground state has only CDW-type insulating phases 
in the strong-NN-coupling.
In particular, the chemical potential width of all CDW insulators
is $U$ in the atomic limit, and the ground state is a 
$(n_0, 0)$ CDW insulator for $(n_0-1) U < \mu < n_0 U$.
As $t$ increases, the range of $\mu$ about which the ground state 
is insulating decreases, and the CDW insulators disappear at 
a critical value of $t_c$, beyond which the system becomes a SS. 
Note that, except for the weak-NN-coupling phase diagram, all of 
the $t_c$ values are comparable to each other for a fixed $zV$.

The main result of this paper is shown in Figs.~\ref{fig:pd}(c-f). 
It is clearly seen that the SS phase occupies a much larger region 
when $zV \gtrsim 1.1U$, existing up to very large hopping values 
which could be orders of magnitude higher than that of the 
weak-NN-coupling. This result is intuitive given that 
the ground state has only $(n_0, 0)$ CDW insulators 
whose sizes are comparable to each other, and that the 
CDW modulations become stronger as $n_0$ (or $\mu$) increases. 
In fact, the SS-SF phase boundary becomes linear in $\mu$ 
and $zt$ when $zV \gtrsim 1.5U$. Our numerical calculations suggest 
that the slope of this line is approximately given by
$
d \mu / d(z t)= 2/(zV/U-1).
$
A similar linear dependence between the particle density and hopping
has recently been found for the SS-SF phase transition boundary 
in the case of quantum rotor model~\cite{buhler}, when the average 
occupation is much higher than unity.

\textit{Experimental realization}:
Here, we argue that our results for two dimensions ($z = 4$) is 
directly applicable to the quasi-two-dimensional dipolar Bose gases. 
For the optical lattice potential
$
V_{\textrm{OL}}(r) = V_0 [\sin^2(kx) + \sin^2(k y)],
$
where $k = 2\pi/\lambda$ is the wave vector and $\ell = \lambda/2$ the
lattice spacing, the on-site interaction depends on the $s$-wave scattering
length $a_s$ via~\cite{bloch}
$
U = \sqrt{8/\pi} k a_s E_r s^{3/4}.
$
Typically, $s = V_0/E_r \sim 10$, where $E_r = \hbar^2 k^2/(2m)$ 
is the recoil energy and $m$ is the particle mass.
Assuming all of the dipoles are polarized along the $z$ direction, 
the dipole-dipole interaction becomes isotropic, leading to
$
V = C_{\textrm{dd}}/(4\pi\ell^3)
$
for the NN repulsion, where $C_{\textrm{dd}} = \mu_0 p^2$ (or 
$p^2/\varepsilon_0$) for particles with permanent magnetic 
(or electric) dipole moment $p$. The ratio
$
U/(4V) = \pi^2\sqrt{2\pi}s^{3/4}/(12\varepsilon_{\textrm{dd}}),
$
where $\varepsilon_{\textrm{dd}} = a_{\textrm{dd}}/a_s$ and
$a_{\textrm{dd}} = m C_{\textrm{dd}}/(12\pi\hbar^2)$ is the dipolar 
length scale, determines the critical threshold ($4V > U$).
For instance, $p = 6\mu_B$ and $a_{\textrm{dd}} = 16a_0$ for the 
$^{52}$Cr atoms~\cite{lahaye}, where $\mu_B$ ($a_0$) is the Bohr 
magneton (radius), and the condition is $a_s \lesssim 8a_0/s^{3/4}$. 
However, $p = 0.6$ Debye and $a_{\textrm{dd}} = 2\times10^3 a_0$ for 
the KRb molecules~\cite{lahaye}, and the condition is 
$a_s \lesssim 10^3a_0/s^{3/4}$. Since $a_s \sim 100a_0$ for Cr atoms,
$a_s$ needs to be tuned via a Feshbach resonance~\cite{lahaye07} 
in order to achieve the critical threshold.

In addition, we note that it is easy to extract the finite-size effects 
of an external trapping potential (e.g. present in atomic systems) 
from Fig.~\ref{fig:pd}. For instance, within the local-density 
approximation, if the center of the trap is a ($2, 0$) CDW indulator 
(say $zt \lesssim 0.6U$), the system is expected to go through first 
a SS, then a ($1, 0$) CDW indulator, before becoming a SF as a function 
of the radial distance towards the edge of the trap. On the other hand, 
if the center of the trap is a SS (say $zt \gtrsim 0.6U$), 
the system is expected to become a SF beyond a critical radius, 
without any intermediate phase.

\textit{Conclusions}:
To summarize, beyond the critical threshold $zV > U$, we showed 
that the SS phase occupies a much larger region in the phase diagram, 
existing up to very large hopping values of the order $zt \gtrsim U$, 
and that the SS-SF phase boundary increases linearly as a function 
of $t$ when $zV \gtrsim 1.5U$. Therefore, our results suggest that 
the prospects of observing the SS phase is much higher when $zV > U$,
which could be easily achieved with quasi-two-dimensional dipolar Bose gases
loaded into optical lattices, by tuning the $s$-wave scattering length 
via currently available Feshbach techniques~\cite{lahaye07}. We believe our mean-field
treatment captures the qualitative physics right, and that this work will 
motivate further quantum Monte Carlo calculations in the strong-NN-coupling 
regime for more accurate phase diagrams.

The author thanks J. K. Freericks for comments.
This work is financially supported by the Marie Curie International 
Reintegration (Grant No. FP7-PEOPLEIRG-2010-268239) and the 
Scientific and Technological Research Council of Turkey 
(Career Grant No. T\"{U}B$\dot{\mathrm{I}}$TAK-3501-110T839).


\begin{thebibliography}{99}
\bibitem{sublattice} This is in contrast to the weak-NN-coupling, where the 
particle and hole excitation energies are higher on the sublattice with higher occupancy,
i.e. the A sublattice, leading to a crystalline (superfluid) order primarily on the sublattice A (B).
\end{thebibliography}
\end{document}